\documentclass[twocolumn]{aastex631}

\usepackage{bm}
\newcommand{\vect}[1]{\bm{#1}}

\shorttitle{UDGs in EAGLE}
\shortauthors{Zheng et al.}

\begin{document}

\title{Ultra-diffuse galaxies in the EAGLE simulation} 

\author[0000-0002-1665-5138]{Haonan Zheng}
\affiliation{Kavli Institute for Astronomy and Astrophysics, Peking University, Beijing 100871, China}
\affiliation{Key Laboratory for Computational Astrophysics, National Astronomical Observatories, Chinese Academy of Sciences, Beijing 100101, China}
\affiliation{Institute for Computational Cosmology, Department of Physics, University of Durham, South Road, Durham, DH1 3LE, UK}

\author[0000-0001-7075-6098]{Shihong Liao}
\affiliation{Key Laboratory for Computational Astrophysics, National Astronomical Observatories, Chinese Academy of Sciences, Beijing 100101, China}

\author[0009-0006-3885-9728]{Liang Gao}
\affiliation{Institute for Frontiers in Astronomy and Astrophysics, Beijing Normal University, Beijing 102206, China}
\affiliation{School of Physics and Microelectronics, Zhengzhou University, Zhengzhou 450001, China}

\author[0000-0001-6115-0633]{Fangzhou Jiang}
\affiliation{Kavli Institute for Astronomy and Astrophysics, Peking University, Beijing 100871, China}

\correspondingauthor{Shihong Liao, Liang Gao, Fangzhou Jiang} \email{shliao@nao.cas.cn, lgao@bao.ac.cn, fangzhou.jiang@pku.edu.cn}

% Abstract of the paper
\begin{abstract}
We use the highest-resolution EAGLE simulation, Recal-L025N0752, to study the properties and formation of ultra-diffuse galaxies (UDGs). We identify 181 UDGs at $z=0$ and find their properties are consistent with observations. The total masses of EAGLE UDGs range from ${\sim}5\times 10^{8}~M_{\odot}$ to ${\sim}2\times 10^{11}~M_{\odot}$, indicating that they are dwarf galaxies rather than failed $L_\star$ galaxies. EAGLE UDGs are not a distinct population, but rather a subset of dwarf galaxies, as their properties generally form a continuous distribution with those of normal dwarf galaxies. Unlike most of the situations in previous studies, the extended sizes of field UDGs in EAGLE are not driven by high halo spins or by supernova-induced stellar expansion, but instead largely arise from high spins in their star-forming gas and thus the newly formed stars at large radii. 
We hypothesize that this might be attributed to galactic fountains, by which star-forming gas is launched to large halo-centric distances and acquires additional angular momentum through interactions with the circumgalactic medium.
For satellite UDGs, ${\sim} 60 \%$ of them were already UDGs before falling into the host galaxy, while the remaining ${\sim} 40\%$ were normal galaxies prior to infall and subsequently transformed into UDGs due to tidal effects after infall.

\end{abstract}


\section{Introduction}
\label{sec:intro}

Ultra-diffuse galaxies (UDGs) are a large sample of low surface brightness galaxies with stellar masses comparable to dwarf galaxies but effective radii as extended as $L_\star$ galaxies \citep{2015ApJ...798L..45V}. They have been observed across different environments, ranging from galaxy clusters to the field \citep[e.g.,][]{2015ApJ...798L..45V,Koda2015,Martinez-Delgado2016,Leisman2017,van_der_Burg2017,Hu2023,Karunakaran2023}. However, there is still no consensus on their formation mechanisms.

Galaxy formation simulations have played a crucial role in revealing the possible origins of UDGs. Although all simulations thus far indicate that UDGs are dwarf galaxies rather than $L_\star$ galaxies, the reasons these dwarf galaxies exhibit such extended sizes remain controversial. Several formation mechanisms have been proposed from different simulation groups:

{\it High halo spins.} Motivated by the classical formation model of galactic discs \citep{Dalcanton1997, Mo1998}, which posits a positive correlation between the effective radius of a disc galaxy and the spin parameter of its host dark matter halo, one scenario proposes that the extended sizes of UDGs originate from the high-spin dark matter halos that they reside in \citep{Amorisco2016}. This scenario is supported by the work of \citet{Rong2017}, which, based on the \citet{Guo2011} semi-analytical model, demonstrates that UDGs form as a result of the late formation times and high spins of their host halos. In addition, \citet{Liao2019} further support this scenario by studying field UDGs in the Auriga simulations \citep{Grand2017,Grand2024}. They find that Auriga UDGs in the field environment preferentially reside in dark matter halos with higher spins compared to normal galaxies. See also \citet{Benavides2023,Benavides2024}, who reach similar conclusions using the TNG50 simulation \citep{Nelson2019,Pillepich2019}, \citet{Newton2023}, who further use constrained simulations adopting the same recipe as Auriga \citep[i.e., HESTIA,][]{Libeskind2020} to predict the number of UDGs in the local group, and \citet{Nandi2025}, who further study the lineage of UDGs with the TNG50 simulation \citep{Nelson2019,Pillepich2019}. 

{\it Supernova feedback-driven outflows and subsequent stellar expansion.} Using the NIHAO simulations \citep{Wang2015}, \citet{DiCintio2017} suggest that the formation of field UDGs results from supernova feedback-driven gas outflows, which can lead to the expansion of both the dark matter and stellar distributions in dwarf galaxies \citep[see also][]{Freundlich2020}. This scenario is further supported by the FIRE \citep{Chan2018}, Horizon-AGN \citep[][as an initial trigger]{Martin2019}, and NewHorizon \citep{Jackson2021} simulations.

{\it Major mergers.} \citet{Wright2021} investigate the field UDGs from the ROMULUS25 simulation \citep{Tremmel2017} and conclude that they primarily form as products of major mergers that occurred at $z \ga 1$. These major mergers temporarily boost the halo and gas spin, then trigger persistent and asymmetric star formation bursts at the galaxy outskirts. Such redistribution of the star formation areas consequently decreases galaxies' central star formation rates (SFRs) and surface brightnesses, and increases the outer SFRs and galaxy effective radii, leaving imprints of older and redder stellar populations in UDG centres. 

{\it Tidal effects and ram pressure.} Environmental factors, such as tidal effects and ram pressure, have also been shown to play a role in the formation of satellite UDGs. Studies supporting this scenario include \citet{Jiang2019}, who use a galaxy group simulation from \citet{Dutton2015}; \citet{Liao2019}, who use the Auriga simulations \citep{Grand2017,Grand2024}; \citet{Carleton2019}, who use the Illustris-dark simulations \citep{Vogelsberger2014, Nelson2015}; 
\citet{Martin2019}, who use the Horizon-AGN simulation \citep{Dubois2014}; 
\citet{Sales2020}, who use the TNG100 simulations \citep{Nelson2019b}; and \citet{Tremmel2020}, who use the RomulusC simulation \citep{Tremmel2019}. In particular, both \citet{Jiang2019} and \citet{Liao2019} find that among satellite UDGs in galaxy groups or Milky Way-like galaxies, nearly half form as field UDGs before being accreted, while the remainder originate as normal galaxies that later evolve into UDGs due to tidal effects.

The diverse formation scenarios seen in different simulations may result from variations in the subgrid implementations of baryonic physics. At present, there is no consensus on the detailed implementations and parameter choices used in galaxy formation simulations \citep[see][for reviews]{Naab2017,Crain2023}. Therefore, studying UDGs across independent simulations is both necessary and useful, as it can offer further insights into their possible formation mechanisms and the connections between physical processes and numerical implementations.

The Evolution and Assembly of GaLaxies and their Environments \citep[EAGLE,][]{Schaye2015} simulations are one of the most influential cosmological galaxy formation simulations performed in recent years, yet they have not been used to study UDGs\footnote{Note that \citet{Kulier2020} used the EAGLE simulation to study low-surface-brightness galaxies (and found that they tend to inhabit high-spin halos), but explicitly excluded dwarfs with stellar masses below $10^{9.5}~M_\odot$, which is the mass range covered in this work. Also note that \citet{Pfeffer2024} used the E-MOSAICS simulation to study the abundance of globular clusters as satellites of UDGs, but did not focus on UDG themselves. }. In this work, we utilize the highest-resolution EAGLE run to investigate the formation of UDGs.

The paper is organized as follows. Section \ref{sec:simulation} describes the EAGLE simulation details and the identified UDG sample. The properties of the EAGLE UDGs are discussed in Section~\ref{sec:properties}, and the formation mechanisms are examined in Section~\ref{sec:formation}. In Section~\ref{sec:conclusion}, we summarize and conclude our results.

\section{Simulation Details}
\label{sec:simulation} 

\subsection{EAGLE simulations (Recal-L025N0752)}

\begin{figure*}
    \centering 
    \includegraphics[width=1.0\columnwidth]{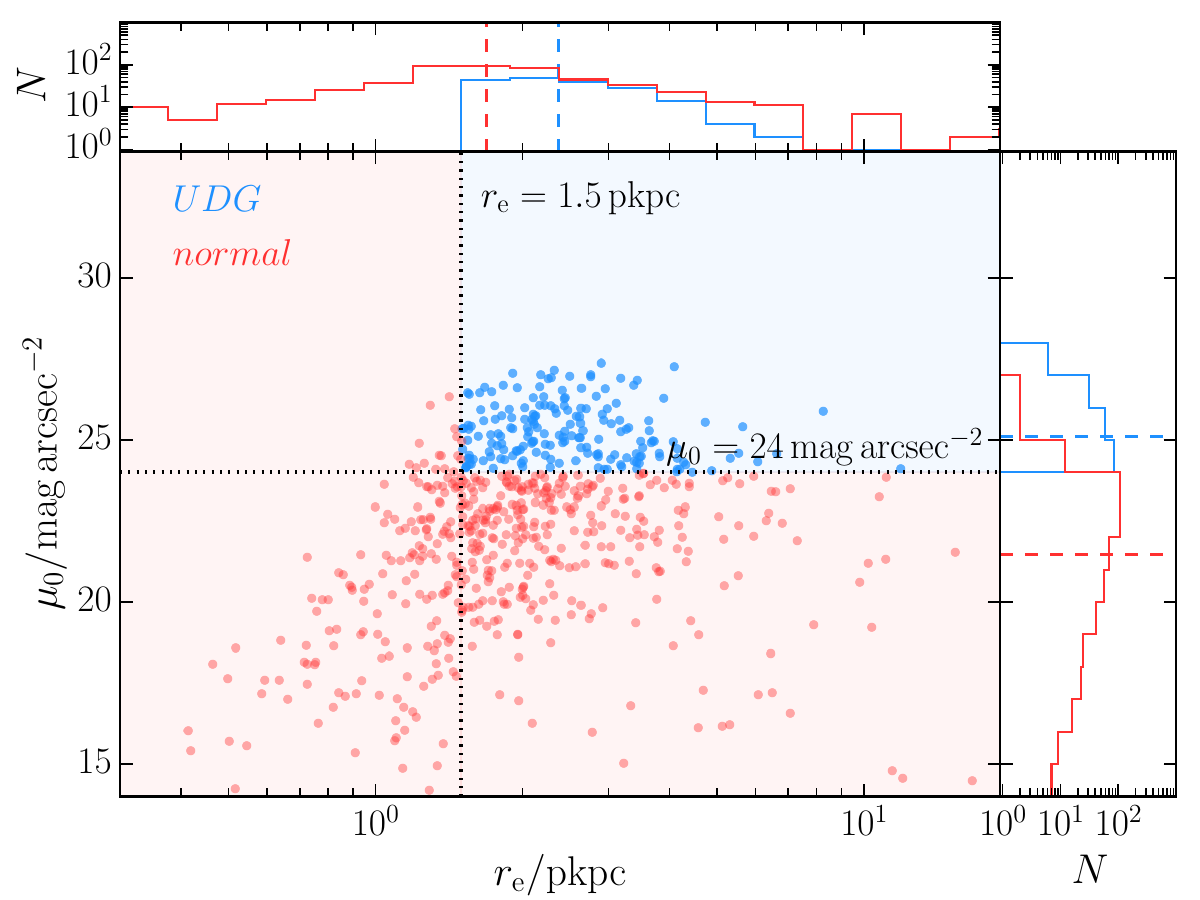}
    \includegraphics[width=1.0\columnwidth]{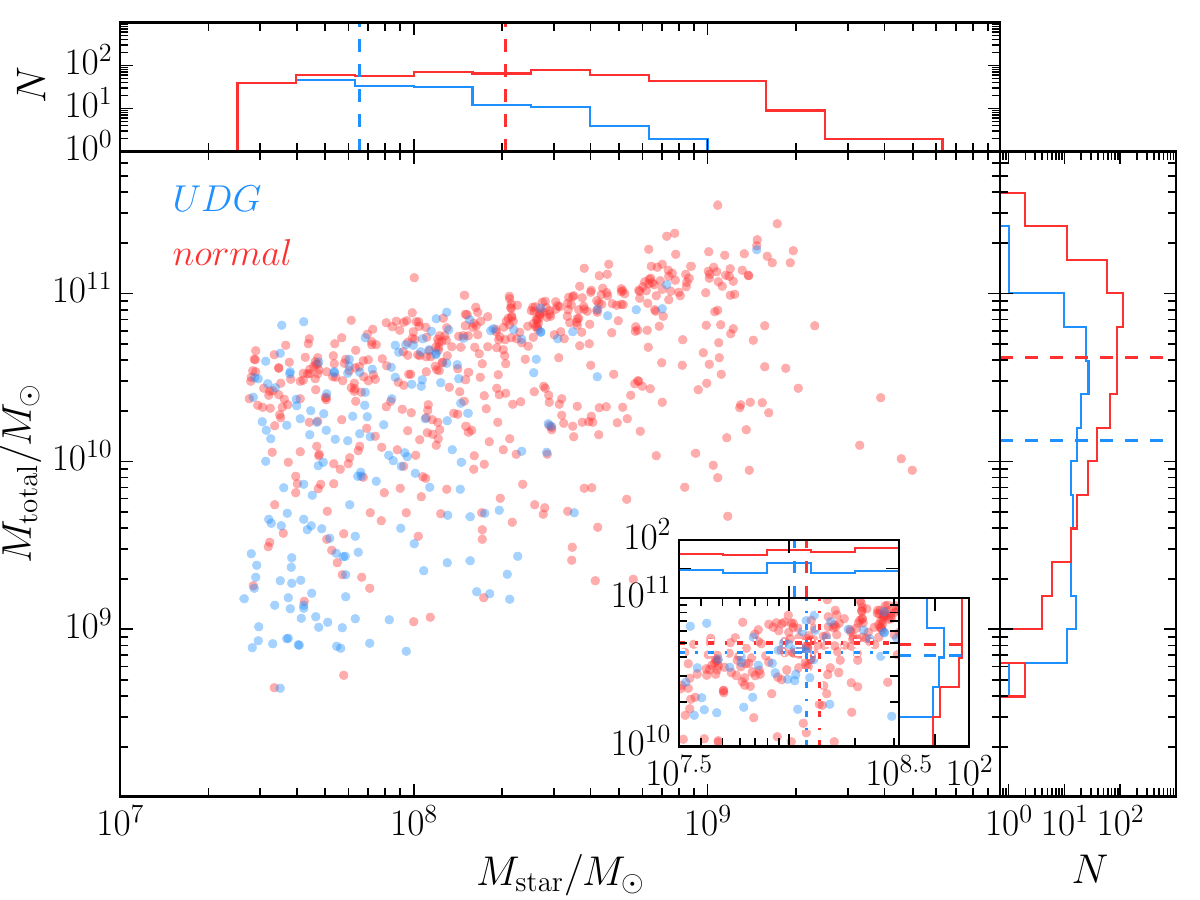}
    
	\vspace{-0.1cm}
	\caption{\textbf{Left}: The relation between the fitted galaxy effective radius, $r_\mathrm{e}$ and central surface brightness, $\mu_0$. UDGs, lying in the blue shaded area of the main panel, are defined as galaxies with $r_\mathrm{e} \geq 1.5~$pkpc and $\mu_0 \geq  24~\mathrm{mag}~\mathrm{arcsec}^{-2}$, while the rest are defined as normal galaxies (red shaded area), as divided by black dotted lines. The upper and right panels show the histograms of $r_\mathrm{e}$ and $\mu_0$, respectively, and the blue (red) dashed lines show the median values of UDGs (normal galaxies). Using this approach, we manage to identify UDGs with median effective radii and central surface brightnesses ($\left<r_{\rm e}\right> = 2.37$ pkpc and $\left<\mu_0 \right> = 25.12$ mag arcsec$^{-2}$) that reasonably match those seen in observations \citep[e.g., $\left<r_{\rm e}\right> = 2.8$ pkpc and $\left<\mu_0 \right> = 25.0$ mag arcsec$^{-2}$ in][]{2015ApJ...798L..45V}. 
    \textbf{Right}: Similar to the left panel, but for the distribution of stellar and total masses. The inset panel displays the properties of field galaxy subsamples selected for comparison (see Section \ref{sec:field} for details), with dash-dotted lines representing their mean stellar and total masses. In the full sample, normal galaxies generally have higher total and stellar masses than UDGs; however, in the subsample, this difference is largely reduced, making it more suitable for direct comparison.}
	\label{fig:fig1}
\end{figure*}

The EAGLE project \citep{Schaye2015} consists of a series of cosmological hydrodynamical simulations with different box sizes and resolutions, performed with a modified version of the \textsc{gadget-3} code \citep[last described in][]{Springel2005}. The EAGLE subgrid model includes a wide range of key galaxy formation processes, such as radiative cooling, reionization, star formation, stellar feedback, metal enrichment, supermassive black hole (SMBH) accretion, and AGN feedback, with details provided in \citet{Crain2015,Schaye2015}. The EAGLE simulations adopt the cosmological parameters from \citet{Planck2014p1}, i.e., $\Omega_{\rm{m}} = 0.307$, $\Omega_{\Lambda} = 0.693$, $\Omega_{\rm{b}} = 0.04825$, $h = 0.6777$, $\sigma_8=0.8288$, and $n_{\rm s} = 0.9611$. The halo catalogue and the particle data from various EAGLE runs are publicly available \citep{McAlpine2016, TheEAGLEteam2017}.  

In this study, we use the highest-resolution EAGLE run, Recal-L025N0752, which initially contains $752^3$ dark matter particles and $752^3$ gas particles in a periodic cubic box with a side length of $25$ cMpc. The initial gas particle mass is $m_\mathrm{g}=2.26\times10^5 ~ M_\odot$, while the dark matter particle mass is $m_\mathrm{dm}=1.21\times10^6 ~ M_\odot$. The comoving gravitational softening length is $1.33 ~ \mathrm{ckpc}$, and it is fixed at $\epsilon = 0.35 ~ \mathrm{pkpc}$ after $z=2.8$. This run adopts the `Recal' parameter set for the subgrid model, which differs from the reference set (`Ref') in the parameters controlling star formation and stellar feedback (i.e., the characteristic density, $n_{\rm H,0} = 0.25$ cm$^{-3}$, and the power-law slope of the density dependence of the thermal energy feedback from star formation, $n_{n} = 1/\ln 10$), the viscosity of the subgrid SMBH accretion disc ($C_{\rm visc} = 2\pi \times 10^3$), and the AGN feedback strength ($\Delta T_{\rm AGN} = 10^{9}$ K). This set of re-calibrated parameters is adopted for high-resolution simulations to improve the match to the observed $z \sim 0$ galaxy stellar mass function. We refer the interested reader to \citet{Schaye2015} for more details.

To identify structures from the EAGLE simulations, dark matter halos are first identified using the friends-of-friends \citep[FoF,][]{Davis1985} algorithm with a linking length parameter of 0.2. Gas and star particles are assigned to the same FoF group as their nearest dark matter particle. Galaxies are then identified as gravitationally bound substructures within FoF groups using the \textsc{subfind} algorithm \citep{Springel2001, Dolag2009}. As defined by these methods, in this paper, we refer to the virial radius and virial mass of a host galaxy in each FoF group ($R_\mathrm{200,\,host}$, $M_\mathrm{200,\,host}$) as the radius within which the mean matter density is 200 times the cosmic critical density and the corresponding enclosed mass, and refer to the total mass of each galaxy $M_\mathrm{total}$ as the total mass of each \textsc{subfind}-identified substructure. 
The halo merger tree is constructed using the \textsc{D-Trees} algorithm \citep{Jiang2014}, and in this work, we mainly use the 17 snapshots after $z=3$ with an average time interval of $\sim 0.7~\mathrm{Gyr}$. 

\begin{figure*}
    \centering
	\includegraphics[width=2.0\columnwidth]{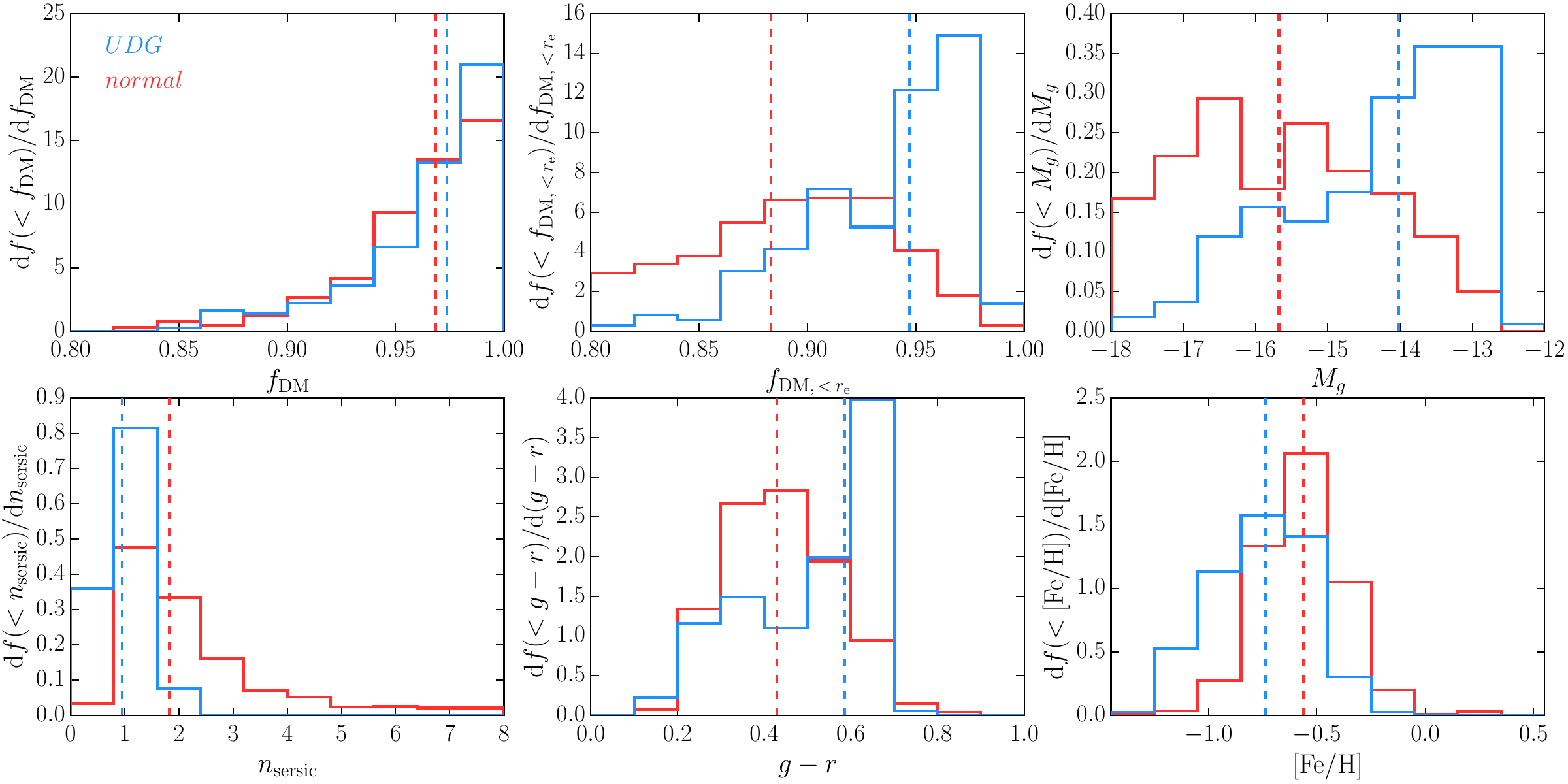}
	\vspace{-0.1cm}
	\caption{General properties of the complete sample of UDGs and normal galaxies at $z=0$. From left to right, the top panels present the PDFs of dark matter fractions of \textsc{subfind} galaxies ($f_{\rm DM}$), dark matter fractions within the fitted galaxy effective radius ($f_{{\rm DM}, <r_{\rm e}}$), and $g$-band magnitude ($M_{g}$), whereas the lower panels show the PDFs of S\'{e}rsic index ($n_{\rm sersic}$), color ($g-r$), and stellar metallicity ($[{\rm Fe}/{\rm H}]$). UDGs and normal galaxies are plotted with blue and red, respectively. The vertical dashed lines mark the median values. Most of the UDGs have a dark matter fraction $\ga$ 90\%, suggesting they are not dark matter deficient galaxies, and they generally tend to have a fainter magnitude, a lower S\'{e}rsic index, a slightly redder color and a lower stellar metallicity, compared to normal galaxies. }
	
	\label{fig:fig3}
\end{figure*}

\subsection{UDG samples}

To identify UDGs from the EAGLE simulations, we follow the method outlined in \citet{Liao2019}. First, each galaxy with at least $N_{\star} = 200$ star particles is rotated to the face-on direction by using the eigenvector associated with the minimum eigenvalue of the inertia tensor. Second, the $g$-band projected surface brightness profile, $I_g (r)$, is computed over the radial range of $r = [3\epsilon, 3 r_{\star, 1/2}]$, divided into $N_{\rm bin} = 10$ logarithmically spaced bins ($r_{\star, 1/2}$ denoting the half-stellar-mass radius)\footnote{This radial range, adopted from \citet{Liao2019}, is a compromise between numerical limitations and reasonable fits. Note that \citet{Liao2019} also tested different radial ranges (e.g., $r = [3\epsilon, 2 r_{\star, 1/2}]$) and different numbers of bins (e.g., 8 and 12) to verify the robustness of the results. }. Here, the luminosity information of the star particles is derived with the method of \citet{Trayford2015}. $I_g (r)$ is then fitted with a \citet{Sersic1963} profile,
\begin{equation}
    I_{g}(r) = I_{\rm e} \exp \left\{-b_n \left[\left( \frac{r}{r_{\rm e}} \right)^{1/n} - 1 \right] \right\},
\end{equation}
where $r_{\rm e}$ is the effective radius, $I_{\rm e}$ is the surface brightness at $r_{\rm e}$, $n$ is the S\'{e}rsic index, and $b_{n}$ satisifies $\gamma(2n, b_n) = \Gamma(2n)/2$, with $\gamma(s, x)$ and $\Gamma(x)$ being the lower incomplete Gamma and Gamma functions respectively. 
The central surface brightness, $\mu_0$, can be converted from $I_g (0)$. We only consider galaxies which return good fits, i.e., with a figure-of-merit function $Q^2 < 0.05$ where $Q^{2} \equiv (1/N_{\rm bin})\sum_{i=1}^{N_{\rm bin}} (\log_{10} I_i^{\rm data} - \log_{10} I_i^{\rm model})^2$.
Finally, we define UDGs as galaxies with $\mu_0 \geq 24~{\rm mag}~{\rm arcsec}^{-2}$, $r_{\rm e} \geq 1.5~{\rm pkpc}$, and $g$-band absolute magnitudes $M_g \in [-18,\ -12]$\footnote{Note that the lower limit on the number of star particles imposes a faintest $g$-band magnitude of ${\sim} -12.5$ on our samples due to resolution limitations. }, following the definition used in the observation work of \citet{2015ApJ...798L..45V}.

\begin{table}
 \caption{Number of UDGs and normal galaxies with $g$-band magnitudes $-18 \leq M_{g} \leq -12$ at $z=0$, classified based on central surface brightness and effective radius as illustrated in Fig. \ref{fig:fig1}. Second column: the total sample number; third/fourth/fifth column: the number and fraction of subsamples further categorized as field/satellite/other (i.e., those not categorized as the former two types) galaxies based on their environment and star formation activities, as described in Section \ref{sec:formation}. }
 \vspace{0.15cm}
 \hspace{-1.35cm}
 \label{tab:table1}
 \begin{tabular}{ccccc}
  \hline
  Type & Total & Field & Satellite & Others \\
  \hline

  UDG & 181 & 63 (35\%) & 73 (40\%) & 45 (25\%) \\   
  Normal & 529 & 336 (64\%) & 110 (21\%) & 83 (16\%) \\ 

  \hline
 \end{tabular}
\end{table}

The left panel of Fig.~\ref{fig:fig1} presents all galaxies from the Recal-L025N0752 run with $-18 \leq M_{g} \leq -12$ and $N_{\star} \geq 200$ (applied at $z=0$) on the $\mu_0$--$r_{\rm e}$ plane. Blue dots represent UDGs which satisfy the aforementioned definition, while red dots indicate normal galaxies that are either more compact ($r_{\rm e} < 1.5$ pkpc) or have brighter centers ($\mu_0 < 24$ mag arcsec$^{-2}$). In total, we identify 181 UDGs and 529 normal galaxies (Table~\ref{tab:table1}), providing a good statistical sample for exploring their properties and formation pathways. The EAGLE UDGs have a median effective radius of $\left<r_{\rm e}\right> = 2.37$ pkpc and a median central surface brightness of $\left<\mu_0 \right> = 25.12$ mag arcsec$^{-2}$, which agree well with the observational values, e.g., 
$\left<r_{\rm e}\right> = 2.8$, 3.9, 1.84 pkpc and $\left<\mu_0 \right> = 25.0$, 24.48, 24.95 mag arcsec$^{-2}$ in \citet{2015ApJ...798L..45V}, \citet{Leisman2017} and \citet{Marleau2021} respectively\footnote{Note that these measurements are carried out in different environments: \citet{2015ApJ...798L..45V} focused on UDGs around the Coma cluster, \citet{Leisman2017} focused on isolated UDGs, and \citet{Marleau2021} looked into low-to-moderate density fields, which may explain the scatters among their results. }. 

\section{UDG properties} \label{sec:properties}

In this section, we study the statistical properties of the EAGLE UDGs. We first compare the stellar ($M_{\rm star}$) and total ($M_{\rm total}$) masses of UDGs and normal galaxies in the right panel of Fig. \ref{fig:fig1}. The stellar masses of UDGs in our sample range from ${\sim} 3 \times  10^{7}~M_\odot$ to ${\sim} 10^{9}~M_\odot$, with a median of $6.53 \times 10^{7}~M_\odot$, which is comparable to the median stellar masses of $\sim 6 \times 10^7~M_\odot$  and $5.13 \times 10^7~M_\odot$ observed in UDGs from the Coma cluster \citep{2015ApJ...798L..45V} and the MATLAS low-to-moderate density fields \citep{Marleau2021} respectively. The total masses of EAGLE UDGs range from ${\sim}5 \times 10^{8}~M_\odot$ to ${\sim}2\times 10^{11}~M_\odot$, with a median of ${\sim}1.5 \times 10^{10}~M_\odot$. This suggests that UDGs in the EAGLE simulations are consistent with dwarf galaxies rather than failed $L_{\star}$ galaxies. 
The stellar and total masses of normal galaxies are higher than those of UDGs on average in the same absolute magnitude range [$-18$, $-12$].

\begin{figure}
    \centering
	\includegraphics[width=1.0\columnwidth]{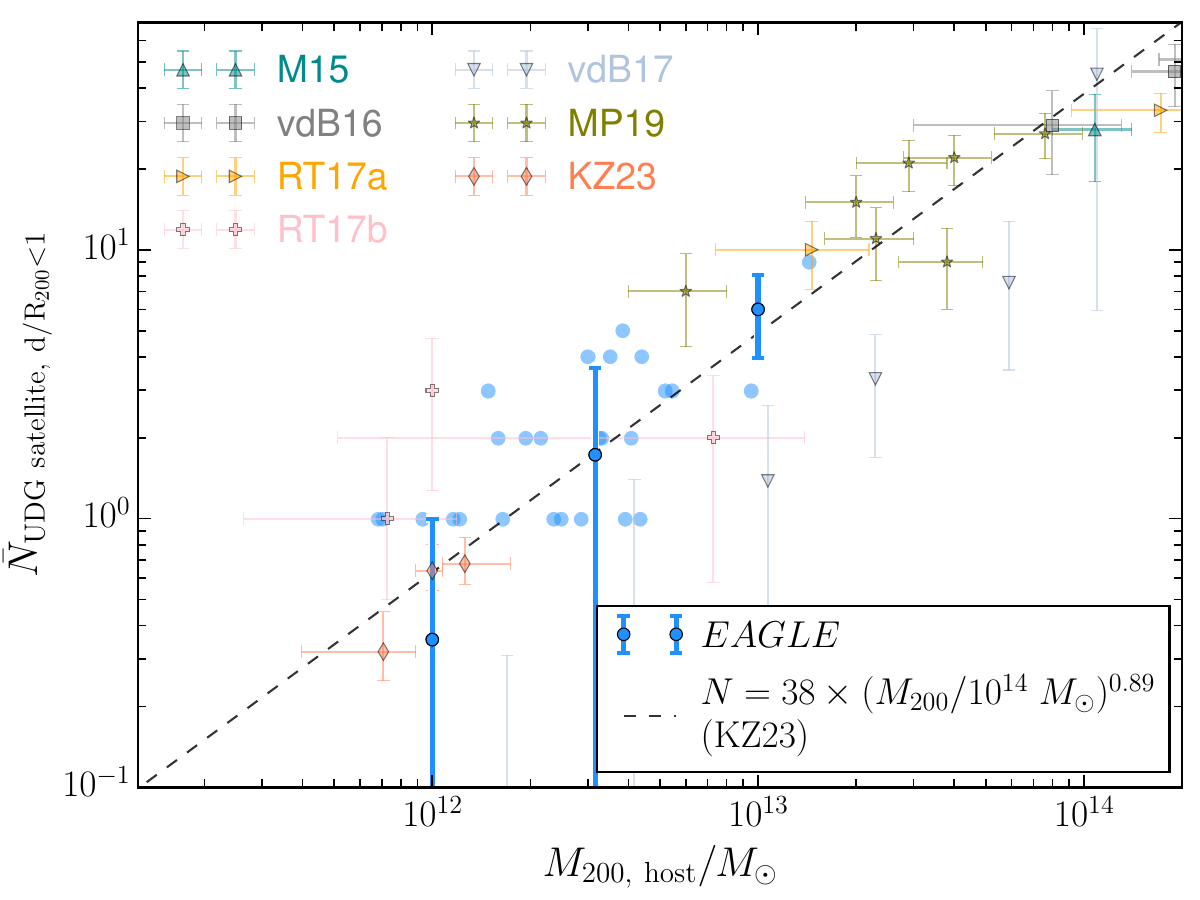}
	\vspace{-0.55cm}
	\caption{The mean UDG satellite abundance as a function of host halo mass, i.e., the $N_\mathrm{UDG\ satellite}-M_\mathrm{200,\,host}$ relation. The individual EAGLE host galaxies are plotted using blue dots, and the mean $N_\mathrm{UDG\ satellite}-M_\mathrm{200,\,host}$ relation computed in different mass bins is shown using blue points with $1\sigma$ error bars. For comparison, we also plot the observational data from \citet[][M15]{Munoz2015}, \citet[][vdB16]{vanderBurg2016}, 
    \citet[][RT17a]{Roman2017a}, \citet[][RT17b]{Roman2017b}, \citet[][vdB17]{van_der_Burg2017}, 
    \citet[][MP19]{Mancera_Pina2019}, 
    and \citet[][KZ23]{Karunakaran2023}. The dashed line shows the best-fitting relation from \citet{Karunakaran2023}. The UDG satellite number from the EAGLE simulation broadly aligns with these observation results at the low mass end. }
	\label{fig:fig4}
\end{figure}

In Fig.~\ref{fig:fig3}, we compare the distributions of several general properties of UDGs and normal galaxies. Observations suggest that a few UDGs may lack dark matter \citep[e.g.,][]{vanDokkum2018,vanDokkum2019}. To investigate whether any EAGLE UDGs are dark-matter-deficient, we compute the dark matter fractions for the entire \textsc{subfind}-identified substructure ($f_{\rm DM}$) and within the fitted galaxy effective radius ($f_{{\rm DM}, <r_{\rm e}}$). Their probability distribution functions (PDFs) are shown in the left and middle panels of the top row. UDGs and normal galaxies exhibit similar distributions of $f_{\rm DM}$. However, UDGs tend to have higher $f_{{\rm DM}, <r_{\rm e}}$ than normal galaxies, likely due to their larger effective radii and more extended baryon distribution\footnote{We also tested with the dark matter fraction within $1\,\mathrm{kpc}$, and the field subsamples as selected in Fig. \ref{fig:fig5} later on – these are consistent with our interpretation here. }. Both $f_{\rm DM}$ and $f_{{\rm DM}, <r_{\rm e}}$ for UDGs and normal galaxies exceed $0.5$, indicating that all UDGs in our sample are dark matter dominated\footnote{We note that dark-matter-deficient galaxies do exist in EAGLE runs, where tidal interactions play a key role in their formation \citep[][with different EAGLE runs: Ref-L0100N1504, Ref-L0025N0752]{Jing2019}, but they are very rare samples ($\sim$ 3.2\% of satellite galaxies in halos with $M_{200}>10^{13}\,M_\odot$) that may not overlap with our UDG population. 
}.

The right panel of the top row presents the PDFs of $g$-band magnitudes, where UDGs are generally fainter, with a median $M_{g} \sim -14$, compared to $M_{g} \sim -15.7$ for normal galaxies. The left panel of the bottom row displays the PDFs of S\'{e}rsic index ($n_{\rm sersic}$), showing that UDGs have a median $n_{\rm sersic} \sim 1$, consistent with the observational results; for example, UDGs in the Coma cluster have a median $n_{\rm sersic} = 0.89$ \citep{Yagi2016}, and those in the MATLAS survey (59 UDGs in low-to-moderate density environments) have a median value of 0.95, obtained from spectral energy distribution fittings \citep{Buzzo2024}. Normal galaxies have a higher median S\'{e}rsic index, $n_{\rm sersic} \sim 2$ and a distribution that extends to larger values. The middle and right panels of the bottom row summarize the distributions of galaxy colors and stellar metallicities\footnote{Stellar metallicities and colors are calculated for all stellar particles in the corresponding subhalos; note that dust-reddening is not included as we intend to study the intrinsic properties of galaxies. }, respectively. Overall, UDGs have distributions similar to those of normal galaxies, with slightly redder colors and lower metallicities.

Fig.~\ref{fig:fig4} shows the relation between satellite UDG abundance and host galaxy mass, comparing it with results from various observations. Here, satellite UDGs are defined as UDGs residing within the virial radius of their host galaxy, $R_{\rm 200, host}$. 
The light blue dots represent the number of satellite UDGs associated with different EAGLE host galaxies, while the heavy blue dots with error bars indicate the mean and standard deviation of satellite UDG abundance in three different mass bins. When computing the mean numbers, we also account for the host galaxies that contain zero UDGs in the corresponding mass bin. 
The EAGLE simulations successfully reproduce the observed UDG abundance\footnote{Note that the UDG definition and volume coverage vary among our listed studies. For example, \citet{Roman2017a, Roman2017b} select UDGs with the central surface brightness as our work does (although \citealt{Roman2017b} adopts a different threshold), while the other works cited here use the average surface brightness within the effective radius, which can cause variations in the measurement of satellite UDG number. } in the Milky Way-sized and group-sized host galaxies. Due to the relatively small box size, no cluster-sized galaxies are identified in the Recal-L025N0752 run. Overall, the EAGLE UDG abundance agrees well with the observational fitting relation from \citet{Karunakaran2023}, given by $N=38 \times (M_{200}/10^{14}\,M_\odot)^{0.89}$.

To conclude this section, our comparison of UDG and normal galaxy properties in the EAGLE simulations suggests that UDGs are not a distinct population but rather a subset of dwarf galaxies, which is consistent with findings from observations and other simulations \citep[e.g.,][]{vanderBurg2016, DiCintio2017, Rong2017, Liao2019, Mancera_Pina2019, Martin2019, Wright2021, Zoller2024, Motiwala2025}. Additionally, the EAGLE simulations successfully reproduce the observed abundance of satellite UDGs. This good agreement between simulations and observations enables us to further investigate the possible formation mechanisms of UDGs.

\section{Formation mechanisms} 
\label{sec:formation}

Following previous studies \citep[e.g.,][]{Liao2019,Jiang2019,Tremmel2020,Wright2021}, we separate the EAGLE UDG sample into field UDGs and satellite UDGs to independently study their formation mechanisms. As defined in Section~\ref{sec:properties}, satellite galaxies are those within the virial radius of their host galaxies, i.e., those with $d/R_{\rm 200, host} \leq 1$ where $d$ is the distance between the centers of a galaxy and its host. 

In contrast, defining field galaxies requires more care, as we cannot simply use the distance criterion $d/R_{\rm 200, host} > 1$. Some galaxies located outside the virial radius of a more massive galaxy at $z=0$ may be backsplash galaxies, i.e., former satellites that have interacted with their host and experienced tidal effects \citep{Liao2019,Benavides2021}. These backsplash galaxies are typically quiescent at $z=0$ and exhibit properties more similar to those of satellites. Including them in the field galaxy sample could contaminate the analysis. Therefore, to ensure a purer field galaxy sample, we apply two criteria: (i) the galaxy must reside in the host halo of a FOF group (i.e., `central' galaxies) or be at least 1.5 times the host halo’s virial radius away from it ($d/R_{\rm 200, host} > 1.5$); (ii) it must have a specific star formation rate (sSFR) higher than $10^{-11}\, \mathrm{yr}^{-1}$.\footnote{We also tested other values (e.g., $10^{-10}$ and $10^{-10.5}\,\mathrm{yr}^{-1}$), and confirmed our conclusions are insensitive to this choice. } Here, ${\rm sSFR}$ denotes the specific star formation rate, defined as ${\rm sSFR} \equiv {\rm SFR}/M_{\rm star}$.

\begin{figure}
    \centering
	\includegraphics[width=1.0\columnwidth]{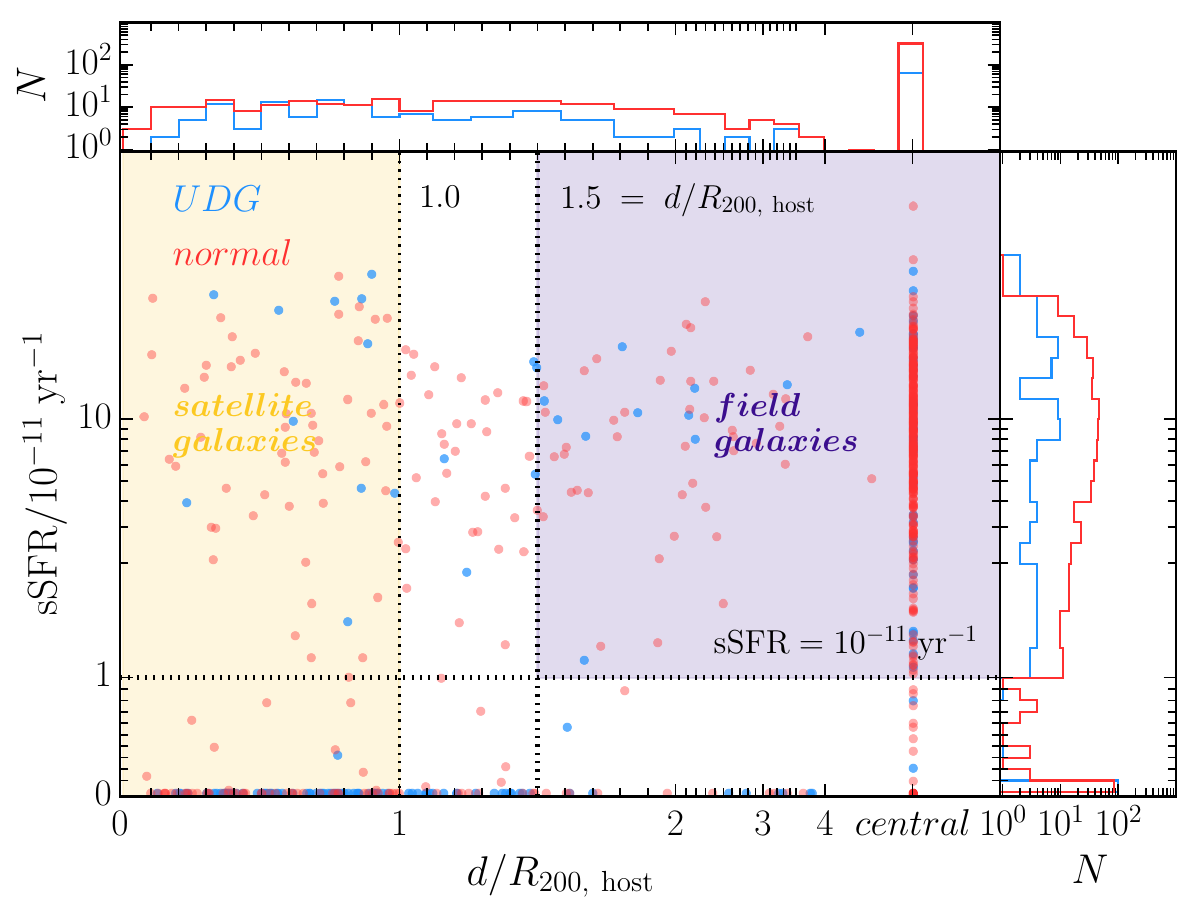}
	\vspace{-0.45cm}
	\caption{The relation between sSFR and relative distance to the host halo. The purple-shaded region in the upper right corner illustrates the criteria for selecting field galaxies (i.e., sSFR $> 10^{-11}~\mathrm{yr}^{-1}$, either as host halos themselves (marked as `central') or as satellite subgroups in an FOF with $d/R_\mathrm{200,\ host}>1.5$). The yellow-shaded region shows satellite galaxies ($d/R_\mathrm{200,\ host} \leq 1$). Histograms of relative distance and sSFR are shown in the upper and right panels, respectively. This division allows us to distinguish the different UDG formation mechanisms in different environments (i.e., in the field and dense environment). }
	
	\label{fig:fig5}
\end{figure}

In Fig.~\ref{fig:fig5}, we investigate the sSFR of EAGLE UDGs (blue dots) and normal galaxies (red dots) as a function of their relative distance to the group center, $d/R_{\rm 200, host}$. Note that the central galaxy samples are shown on the right-hand side for clarity and account for most of the field galaxies. Overall, both UDGs and normal galaxies span similar ranges of $d/R_{\rm 200, host}$ and ${\rm sSFR}$, reinforcing the idea that UDGs are an extreme subset of dwarf galaxies rather than a distinct population. The yellow-shaded region marks the satellite galaxies, while the purple-shaded region corresponds to field galaxies, as defined above. This figure also reveals the presence of galaxies outside the virial radius of a more massive galaxy with very low or even zero sSFR, indicating that they might be backsplash galaxies \citep[e.g.,][]{Geha2012}.

\begin{figure*}
    \centering
    \includegraphics[width=2.0\columnwidth]{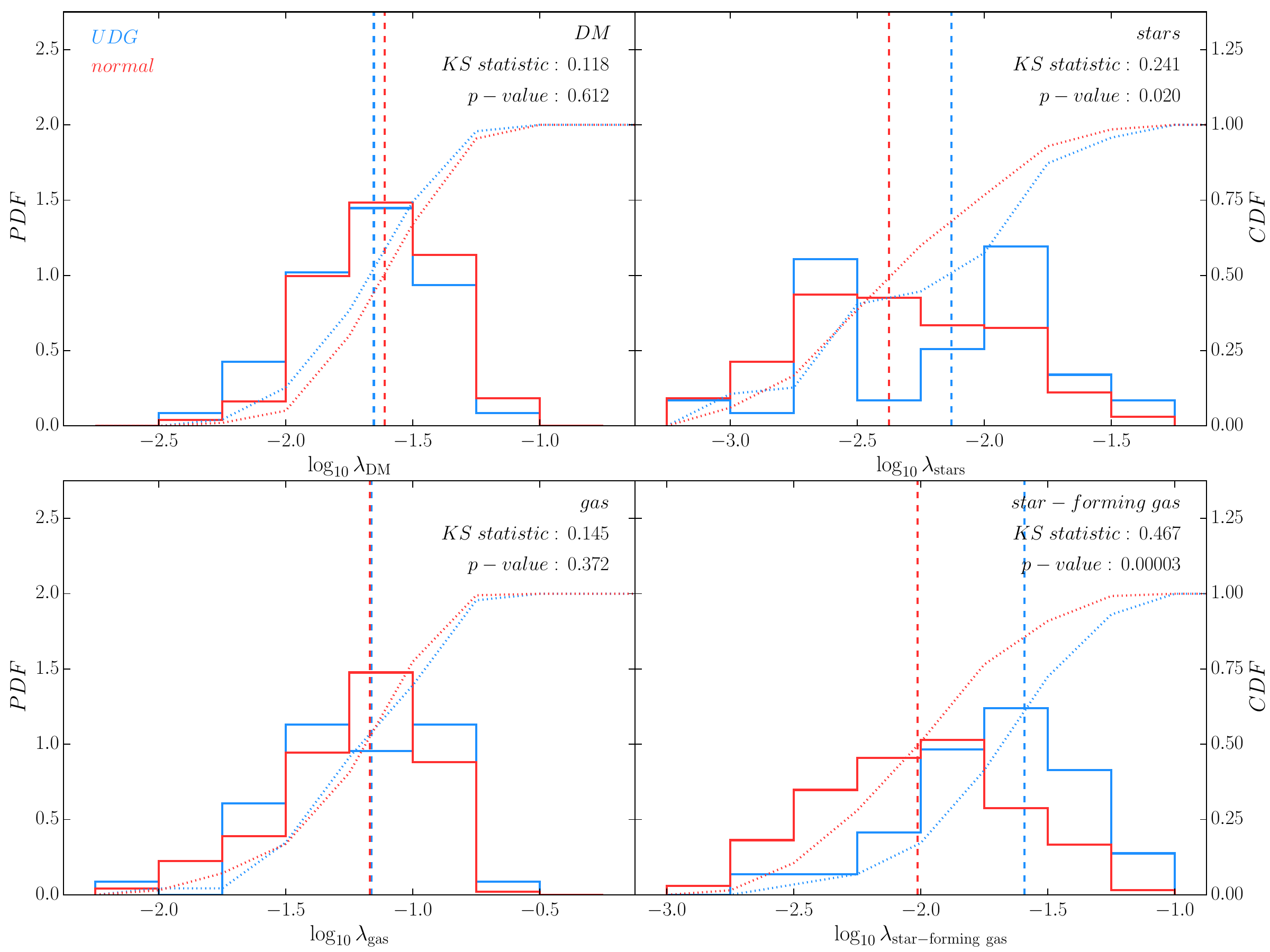}
	\vspace{-0.1cm}
        \caption{PDFs of spin parameters for field galaxies. From left to right, top to bottom, the distributions for dark matter, stars, gas, and star-forming gas (i.e., gas with ${\rm SFR} > 0$) are shown. Blue and red lines represent UDGs and normal galaxies, respectively. The dotted lines show the cumulative distribution functions (CDFs, right $y$-axis). The vertical dashed lines indicate the median spin parameter of each distribution. The KS test results are summarized in the top-right corner of each panel. UDGs have higher spins in star-forming gas and stars than normal galaxies, while the differences in dark matter and gas spins are insignificant. }
	
	\label{fig:fig6}
\end{figure*}

As shown in columns 3 and 4 of Table~\ref{tab:table1}, our sample contains 63 (336) field UDGs (normal galaxies) and 73 (110) satellite UDGs (normal galaxies). This shows that UDGs have a higher occurrence as satellites than in the field and indicates that environment might be a factor worth considering in their formation mechanism. The number of galaxies that do not meet either criteria (i.e., those in the white region of Fig.~\ref{fig:fig5}) is listed in the last column of Table~\ref{tab:table1}. In the following subsections, we investigate the formation mechanisms of field and satellite UDGs separately.

\subsection{UDG formation in the field} \label{sec:field}

In this subsection, we compare the properties of UDGs and normal galaxies in the field, including their spins, formation times, density profiles, and other relevant characteristics, to investigate the formation mechanisms of field UDGs. To ensure high numerical resolution and a fair comparison, we select a field galaxy subsample with similar mass distributions, as shown in the inset panel of Fig. \ref{fig:fig1}. Specifically, we select field galaxies with total masses in the range of [$10^{10}~M_\odot$, $10^{11}~M_\odot$], stellar masses between [$10^{7.5}~M_\odot$, $10^{8.5}~M_\odot$], and a continuously traceable mass accretion history after $z=3$. This relatively tight mass range results in a sample of 47 UDGs and 197 normal galaxies, providing a suitable basis for detailed comparisons. The mean stellar (total) mass of UDGs and normal galaxies in this subsample are $1.06 \times 10^{8} ~ M_\odot$ ($4.09 \times 10^{10} ~ M_\odot$) and $1.21 \times 10^{8} ~ M_\odot$ ($4.86 \times 10^{10} ~ M_\odot$), respectively, making them close enough for comparisons of galaxy spin, density profiles, and other structural properties.

\subsubsection{Spins of different components}
\label{subsec:spin}

To examine whether the EAGLE field UDGs form in high-spin halos, as suggested by previous semi-analytical models and hydrodynamical simulations \citep[e.g.,][]{Amorisco2016,Rong2017,Liao2019,Benavides2023}, we compare the spin parameter distributions of UDGs and normal galaxies. Following \citet{Bullock2001, Peebles1969}, the dimensionless spin parameter for a given component $\alpha$ is defined as
\begin{equation}
    \lambda_{\alpha} \equiv \frac{j_{\alpha}(<R)}{\sqrt{2} R V_{\rm circ}(R)},
\end{equation}
where $R$ is twice the half-total-mass radius of the galaxy, and the circular velocity is given by $V_\mathrm{circ}(R)=\sqrt{GM_{\rm tot}(<R)/R}$, where $G$ is the gravitational constant and $M_{\rm tot}(<R)$ is the total enclosed total mass within $R$. The specific angular momentum of component $\alpha$, $j_{\alpha}(< R)$, is the magnitude of 
\begin{equation}
    \vect{j}_{\alpha} = \frac{\sum_i m_{\alpha, i} \left(\vect{r}_{\alpha, i}-\vect{r}_\mathrm{c}\right) \times \left(\vect{v}_{\alpha, i}-\vect{v}_\mathrm{c}\right)}{\sum_i m_{\alpha, i}}, 
\end{equation}
where $\vect{r}_\mathrm{c}$ and $\vect{v}_\mathrm{c}$ denote the center-of-mass position and velocity of the dwarf galaxy, $m_{\alpha, i}$, $\vect{r}_{\alpha, i}$, $\vect{v}_{\alpha, i}$ represent the mass, position, and velocity of the $i$-th particle of component $\alpha$ (i.e., dark matter, star, gas, and star-forming gas), respectively. The summation is taken over all particles of the component $\alpha$ within $R$.

The spin parameter distributions for different components are shown in Fig. \ref{fig:fig6}. The distributions of dark matter spin are largely similar – the median spin for UDGs (0.022) is very close to that of normal galaxies (0.025). A Kolmogorov–Smirnov (KS) test returns a KS statistic of 0.118 and a $p$-value of 0.612, indicating that these two distributions are very likely drawn from the same parent distribution. The gas spin distributions are also quite close, with median values of 0.069 for UDGs and 0.068 for normal galaxies. The KS test yields a statistic of 0.145 and a $p$-value of 0.372, further confirming the similarity. The difference in stellar spin is slightly larger – the median stellar spin of UDGs (0.007) is $\sim$ 50\% higher than that of normal galaxies (0.004). This is supported by the KS test, which gives a KS statistic of 0.241 and a $p$-value of 0.020. This difference echos the radial distribution of stellar particles, as UDGs generally have larger sizes. Note that in both UDGs and normal galaxies, the gas component exhibits the highest spins, followed by dark matter, while the stellar component has the lowest median spin. This trend is consistent with the results from the Illustris simulations presented in \citet{Zjupa2017}.

Therefore, unlike Auriga and IllustrisTNG simulations, field UDGs in the EAGLE simulation do not reside in higher-spin dark matter halos compared to normal dwarf galaxies. This result aligns with the findings from \citet{Yang2023}, who showed that while a positive correlation between halo spin and galaxy size exists for dwarf disc galaxies in the Auriga simulations (or the so-called Illustris simulation family), such a correlation is not observed in the APOSTLE simulations (or the EAGLE simulation family).

However, compared to normal dwarfs, UDGs do exhibit higher spins in their stellar components, which are likely inherited primarily from the gas that formed the stars. This motivates us to further compare the spins of star-forming gas in UDGs and normal dwarf galaxies.
In the bottom-right panel of Fig.~\ref{fig:fig6}, we select gas particles that are currently undergoing star formation (i.e., with ${\rm SFR} > 0$) and compute their spin parameters as outlined above.\footnote{Note that four normal galaxies with less than 50 star-forming gas particles within twice the half-total-mass radius of the galaxy, which could potentially suffer from resolution effects, are discarded.} We find that the median spin parameter of star-forming gas in UDGs (0.026) is more than twice that of normal galaxies (0.010), aligning with the KS statistic and $p$-value of 0.467 and 0.00003, respectively. 
The physical origin of such higher spins of star and star-forming gas (despite the rather similar distributions of dark matter and total gas spins), will be discussed in Section \ref{sec:discussion}.

\begin{figure}
    \centering
    \includegraphics[width=1.0\columnwidth]{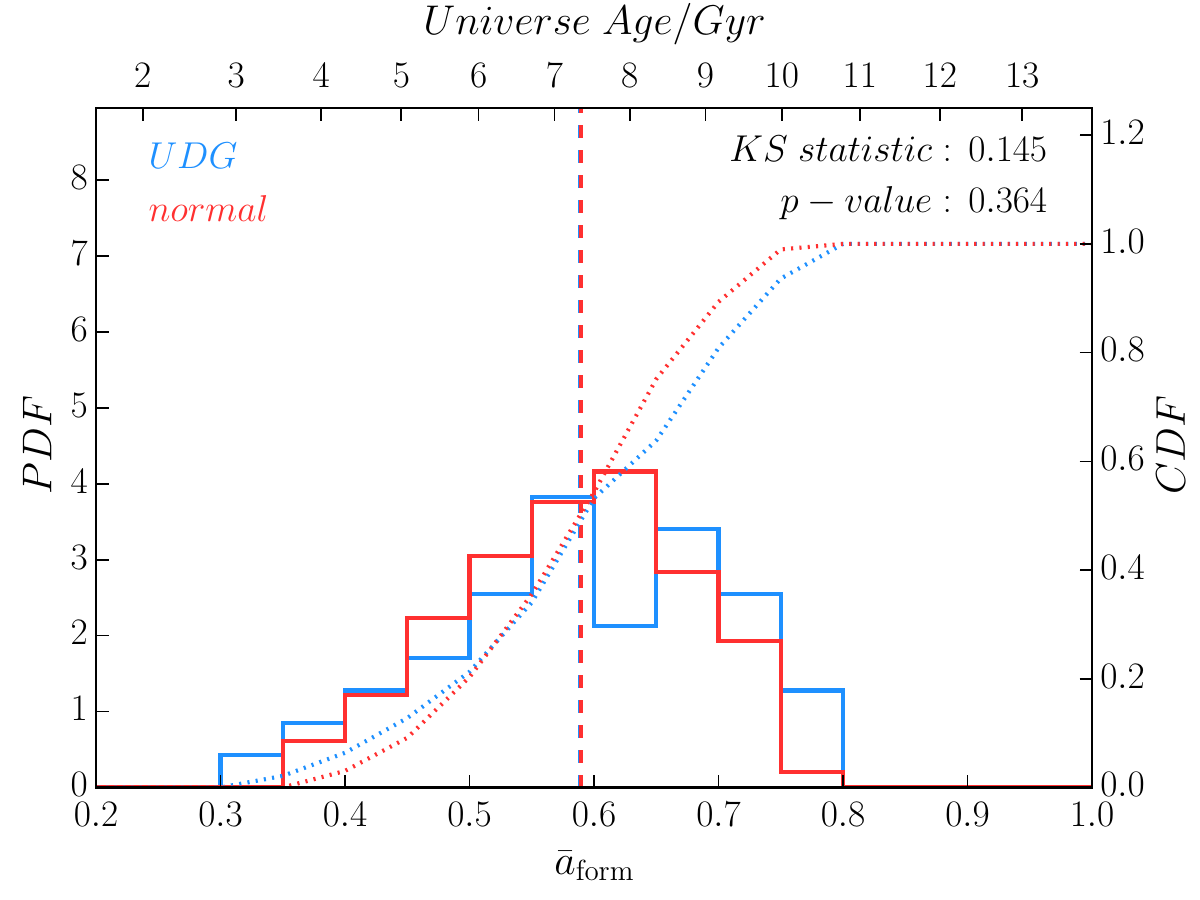}
	\vspace{-0.55cm}
	\caption{Similar to Fig. \ref{fig:fig6}, but for the PDFs (solid, left $y$-axis) and CDFs (dotted, right $y$-axis) of the mean stellar formation time. The stellar population of UDGs is only marginally younger than that of normal galaxies, with a difference of $0.013 ~\mathrm{Gyr}$ in their medians. }
	
	\label{fig:fig7}
\end{figure}

\begin{figure*}
    \centering
	\includegraphics[width=2.0\columnwidth]{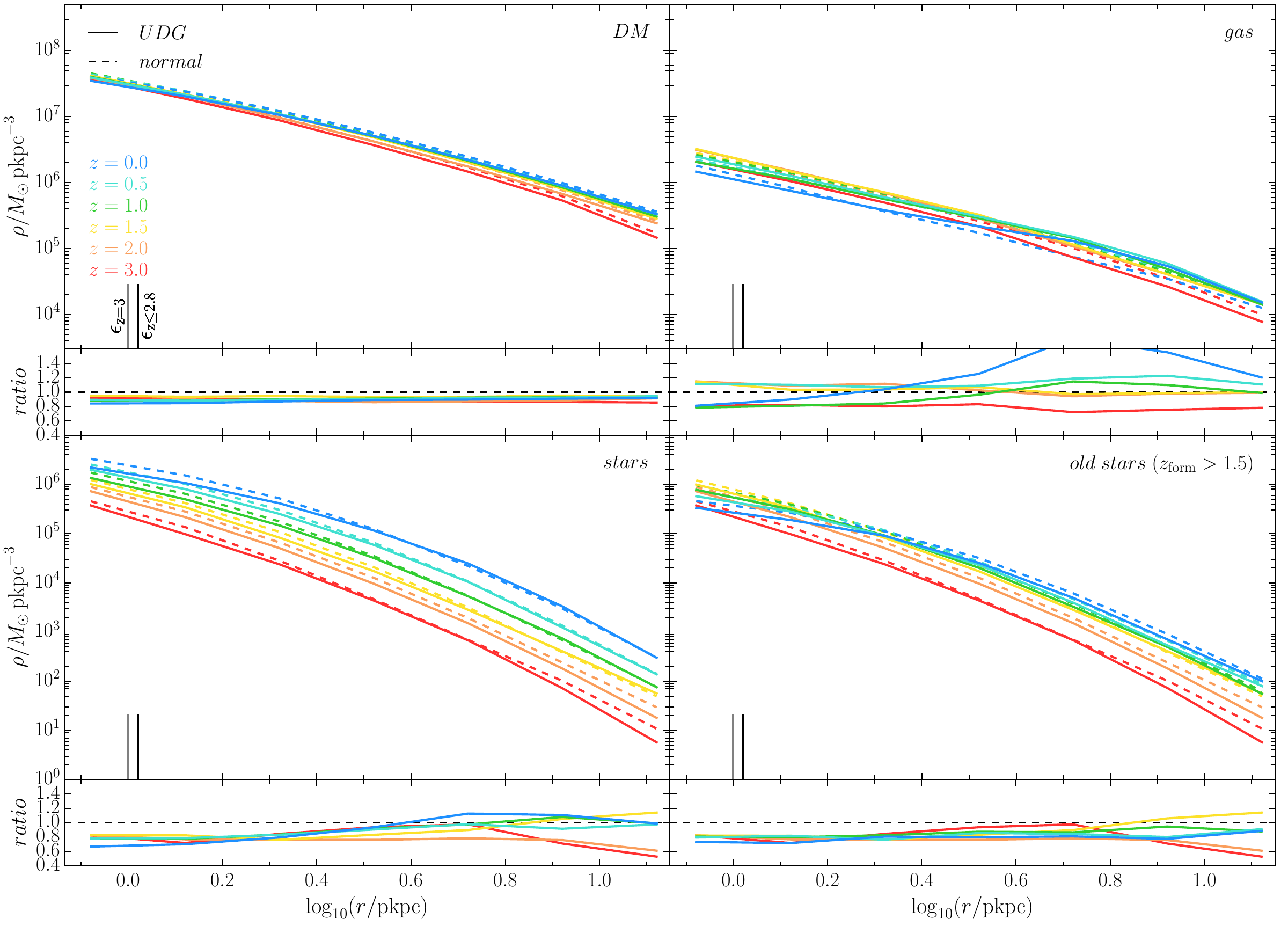} 
	\vspace{-0.15cm}
	\caption{Stacked density profiles of different components in field galaxies with $\log_{10} \left(M_\mathrm{total}/M_\odot\right) \in [10, 11]$ and $\log_{10} \left(M_\mathrm{star}/M_\odot\right) \in [7.5, 8.5]$. UDGs and normal galaxies are represented by solid and dashed lines, and different colors show the profiles of the same halos traced at different redshifts. 
    The vertical line segments in the lower left corner represent three times the gravitational softening length. 
    The bottom subpanels below main panels show the ratio between UDGs and normal galaxies. 
    While the dark matter profiles suggest similar shapes without apparent cores, the profile of gas and stars in UDGs are slightly less concentrated compared to normal galaxies. UDGs have fewer old stars, indicating more recent star formation activities. The profiles of old stars are mildly expanding after $z=1$, with rates similar among UDGs and normal galaxies. 
	}
	\label{fig:fig8}
\end{figure*}

\subsubsection{Formation time of stellar populations}

Using the L-Galaxies semi-analytical model \citep{Guo2011}, \citet{Rong2017} suggested that UDGs form later than normal dwarf galaxies, as objects that form later tend to be more extended due to the expansion and dilution of the Universe. To examine whether EAGLE UDGs exhibit a similar trend, we compare the PDFs of the mean stellar formation time ($\bar{a}_{\rm form}$) for UDGs and normal dwarf galaxies in Fig. \ref{fig:fig7}.

The stellar formation time distributions of UDGs and normal dwarf galaxies show only minimal differences. UDGs form slightly later than normal galaxies, with a median formation time of 7.33 Gyr compared to 7.35 Gyr for normal dwarfs. A KS test gives a KS statistic of 0.145 and a $p$-value of 0.364, confirming that the difference between these two PDFs is statistically insignificant.

\subsubsection{Evolution of density profiles}
\label{subsec:core}
Using the NIHAO simulations, \citet{DiCintio2017} proposed that field UDGs form due to supernovae-driven gas outflows. As gas is expelled from the galaxy center, the resulting fluctuations in the gravitational potential cause both dark matter and stars to expand. This process can be illustrated by examining the time-evolution of stellar and dark matter density profiles. For example, in figure 3 of \citet{DiCintio2017}, the central stellar density keeps decreasing from $z=4$ to $0$ as stars migrate outward, leading to lower central luminosity and extended stellar size. By $z = 1.5$, the galaxy meets the criteria of a UDG. This evolution suggests that in the NIHAO simulations, UDGs primarily form through stellar expansion rather than new star formation in the outskirts. It is worth noting that \citet{Benitez-Llambay2019} studied the dependence of the inner dark matter density profile on the variation of gas density thresholds for star formation, $\rho_\mathrm{th}$, with modifications on the EAGLE recipe. In their figure 2, using zoom-in runs with different $\rho_\mathrm{th}$, they found a threshold of $\rho_\mathrm{th}=0.1~\mathrm{cm}^{-3}$ would result in a cuspy dark matter profile, while relatively high values of $\rho_\mathrm{th}$ ($\sim 10-160~\mathrm{cm}^{-3}$) could possibly induce a dark matter core in their most massive galaxy run ($M_{200}=10^{10.6}~M_\odot$, $M_\mathrm{galaxy}=10^{8.0}~M_\odot$, which lies within the mass range being tested in this paper). The EAGLE run we use here does not adopt a fixed $\rho_\mathrm{th}$ as simplified in \citet{Benitez-Llambay2019}, but is allowed to increase to $10~\mathrm{cm}^{-3}$ (from $0.1~\mathrm{cm}^{-3}$) in metal-poor systems, overlapping with the range where \citet{Benitez-Llambay2019} find core formation possible. Therefore, it is still necessary for us to examine this UDG formation channel associated with dark matter cores, especially for galaxies possibly stemming from metal-poor gas. 

To investigate whether EAGLE field UDGs are driven by feedback mechanisms, we examine the evolution of density profiles for different components (e.g., dark matter, gas, and stars) in Fig.~\ref{fig:fig8}. To reduce noise, we present stacked density profiles, averaging the matter distributions across all 47 UDGs and 197 normal dwarfs in our field subsample.

The top-left panels show the dark matter density profiles for UDGs (solid lines) and normal dwarfs (dashed lines) from $z=3$ to $0$. UDGs and normal dwarfs exhibit similar profiles at different redshifts, though UDGs tend to have densities that are ${\sim}10$--$20\%$ lower than those of normal dwarfs. Notably, both UDGs and normal dwarfs exhibit cuspy dark matter density profiles, in contrast to the cored profiles that emerge at low redshifts due to supernovae-driven feedback in the NIHAO simulations presented by \citet{DiCintio2017}. This suggests that strong feedback-driven stellar expansion may not be a dominant mechanism for UDG formation in EAGLE.

The gas density profiles are presented in the top-right panels. For UDGs, the central gas density ($r \la 2$ pkpc) tends to decrease from $z = 2$ to $0$, while the gas density at outer radii ($r \sim 10$ pkpc) tends to increase as redshift decreases. In contrast, normal galaxies show less evolution in central gas density and less growth in outer gas density compared to UDGs. In UDGs, the accumulation of gas in the outer regions could lead to increased star formation at these locations, as discussed below. The subpanel below shows the ratio between UDGs and normal galaxies. In this subpanel, we find the gas density exhibits more fluctuations than the dark matter density, as gas is influenced by stochastic star formation and stellar feedback effects.

The bottom-left panels show the evolution of stellar density profiles. At $z=4$, the progenitors of UDGs and normal dwarfs exhibit similar stellar density profiles in shape, with UDGs having amplitudes lower by ${\sim}20$--$30\%$. As redshift decreases, stellar density increases steadily across the entire radial range. This differs from the findings of \citet{DiCintio2017}, where central stellar density decreases while outer stellar density enhances due to feedback-driven expansion. By $z=0$, compared to normal dwarf galaxies, EAGLE UDGs experience more stellar growth at large radii, resulting in a more flattened stellar distribution, ultimately leading to their UDG-like nature. Note that this is consistent with the higher spin observed in the star-forming gas, as discussed in the previous subsection.

The bottom-right panels show the evolution of density profiles for old stars, i.e., those formed before $z = 1.5$. After $z = 1.5$, the central stellar density ($r \la 2$ pkpc) of UDGs decreases slightly, while the outer density profile ($r \sim 10$ pkpc) increases mildly, indicating a mild outward migration of stars. However, when compared to the total stellar density profiles in the bottom-left panels, it is evident that the overall profiles are dominated by stars formed after $z=1.5$. This suggests that new star formation plays a more significant role in driving the extended stellar distribution observed at $z=0$.

Compared to the features of dark matter core and stellar expansion proposed by \citet{DiCintio2017}, we find that the formation of EAGLE field UDGs is unlikely to be driven by the redistribution of stars in response to supernova feedback–driven outflows. Instead, EAGLE field UDGs on average tend to have more new stars forming at larger radii which lead to flatter stellar distributions and, consequently, to larger effective radii.

\subsubsection{Discussion}
\label{sec:discussion}

In the previous subsections, we explored a few potential origins for the extended sizes of field UDGs, including high halo spins, late stellar formation times, and stellar expansion driven by supernova feedback. However, none of these mechanisms appear to be strong contributors to the formation of field UDGs in EAGLE. 
Nevertheless, we do observe significantly higher spins in the stellar and star-forming gas components of EAGLE UDGs compared to normal galaxies, along with a more extended gas and stellar distribution after $z \la 1$. 

\begin{figure}
    \centering
    \includegraphics[width=1.0\columnwidth]{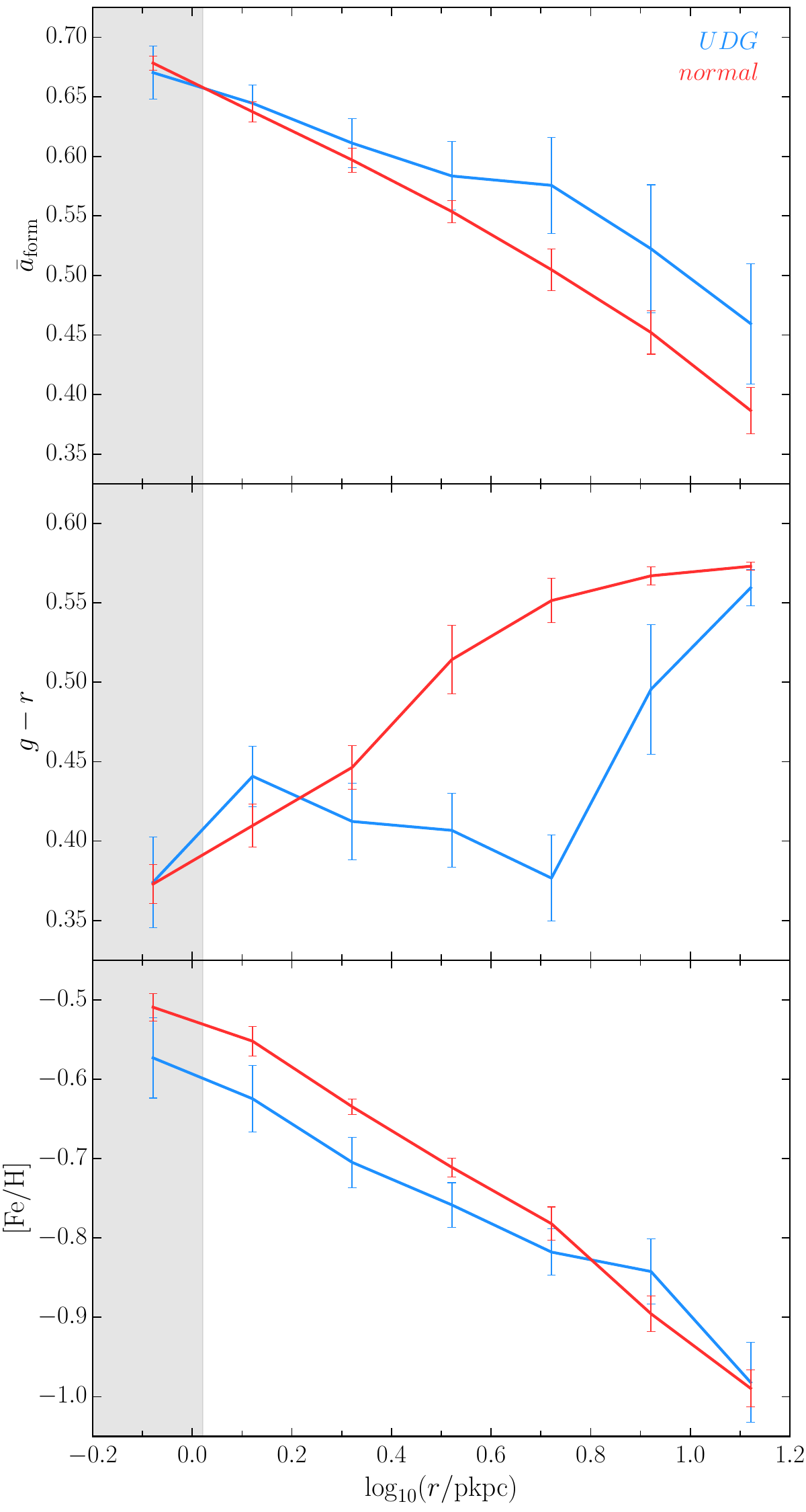}
	\vspace{-0.45cm}
	\caption{Median radial profiles of mean star formation time, color, and stellar metallicity for field galaxies at $z=0$. 
    Error bars show the error of median estimated using bootstrap resampling. The gray shaded regions indicate the radii that are less than 3 times the softening length at $z=0$. Compared to normal galaxies, UDGs tend to have slightly shallower slopes in these profiles, suggesting a connection to gas outflow and recycling processes.} 
	
	\label{fig:fig9}
\end{figure}

A further study on the history of the spun-up star-forming gas requires a compactly traced merger tree, but unfortunately the number of snapshots for the EAGLE run we use here is too limited (i.e., only 10 snapshots after $z=1$) for us to do so. 
Instead, we place our results in context by comparing them with prior studies of UDG formation mechanisms.
For example, \citet{DiCintio2017} found a more extended distribution of HI in the UDGs of the NIHAO simulation, (which aligns with our findings of more extended star-forming regions in UDGs), but as we have addressed in Section \ref{subsec:core}, we do not find a dark matter core or expansion of the stellar component in EAGLE UDGs.  
\citet{Benavides2024} found the UDGs in the TNG50 simulation tend to be more dark matter dominated than normal galaxies (which aligns with our Fig. \ref{fig:fig3}) and have steeper metallicity profiles (which aligns with the Fig. \ref{fig:fig9} we show after), but in Section \ref{subsec:spin} we do not find the higher dark matter spin they suggested for UDG formation.
One of the remaining mechanisms in the literature would be major mergers as proposed by \citet{Wright2021}. 
We cannot directly compare to their analysis (cf. their Figure 14) due to the lack of snapshot outputs to determine the last major merger time, but we examined the evolution of halo spin (not shown) and found the median of those of UDGs is persistently slightly lower than the median of normal galaxies' in the history. 
This indicates that EAGLE UDGs may not originate from the major mergers which typically happen 8-12 Gyrs ago, lift halo spin and redistribute star formation region as proposed in \citet{Wright2021}. 
Additionally, in our later figure (Fig. \ref{fig:fig9}), we found a bluer stellar population in galaxy outskirts, compared to a flatter color gradient in the Romulus25 galaxies, indicating nuance in the overall star formation history between different simulation recipes. 
All in all, these possible mechanisms show some properties similar to the EAGLE UDGs, but suggest other deviating features in the meantime; therefore, we propose a possible UDG origin related to the angular momentum of star-forming gas as follows. 

Recent hydrodynamical simulations and observations \citep[e.g.,][]{Brook2012,DeFelippis2017,Grand2019,Romeo2023,Yang2024} have already shown that the evolution of baryonic angular momentum is more complicated than the classical picture, in which the hot gas has the same specific angular momentum as the host dark matter halo, cools radiatively and forms stars while conserving angular momentum, and ultimately determines the size of the stellar disc \citep[e.g.,][]{Fall1980,Mo1998}. These simulations reveal that after gas cools and enters the disc, part of it is ejected from the high-density interstellar medium (ISM) region, travels to the circumgalactic medium (CGM) region, and later returns to the disc – a process that can occur multiple times for a given gas element. This is known as a galactic fountain \citep[see][for a review]{Fraternali2017}. During this process, the ejected gas can mix with the CGM and acquire angular momentum from it. 
As a result, the angular momentum of the gas that eventually forms stars can differ from that of their host dark matter halo. 

Unfortunately, we are unable to trace the detailed cycles of star-forming gas particles or stellar particles as done in previous studies, due to the limited number of snapshot outputs available for the EAGLE run used in this work (i.e., only 10 snapshots after $z=1$). The mean time interval between snapshots is $\sim 0.7$ Gyr, exceeding the typical recycling timescale \citep[$\sim 0.3-0.5$ Gyr according to][]{Grand2019}. 

Therefore, we refer to the findings in \citet{Yang2024}, which traced the evolution of the angular momentum of star and the corresponding gas particles using APOSTLE--AURIGA simulations \citep{Sawala2016, Fattahi2016,Kelly2022}.
In their Fig. 6, both APOSTLE (which inherits the same subgrid physics from EAGLE) and AURIGA recipe runs show an increase in particle angular momentum between the first and last snapshots when gas particles satisfy the star formation condition, and this increase is significantly stronger for particles that undergo baryon cycling. They also found that galaxies in the AURIGA recipe runs have higher fractions of recycled gas, leading to more significant jumps in the angular momentum of all tracers during the recycling process.\footnote{Note that the final stellar angular momentum in each individual galaxy also depends on the initial gas angular momentum at accretion time. }

\begin{figure*}
    \centering
	\includegraphics[width=1.4\columnwidth]{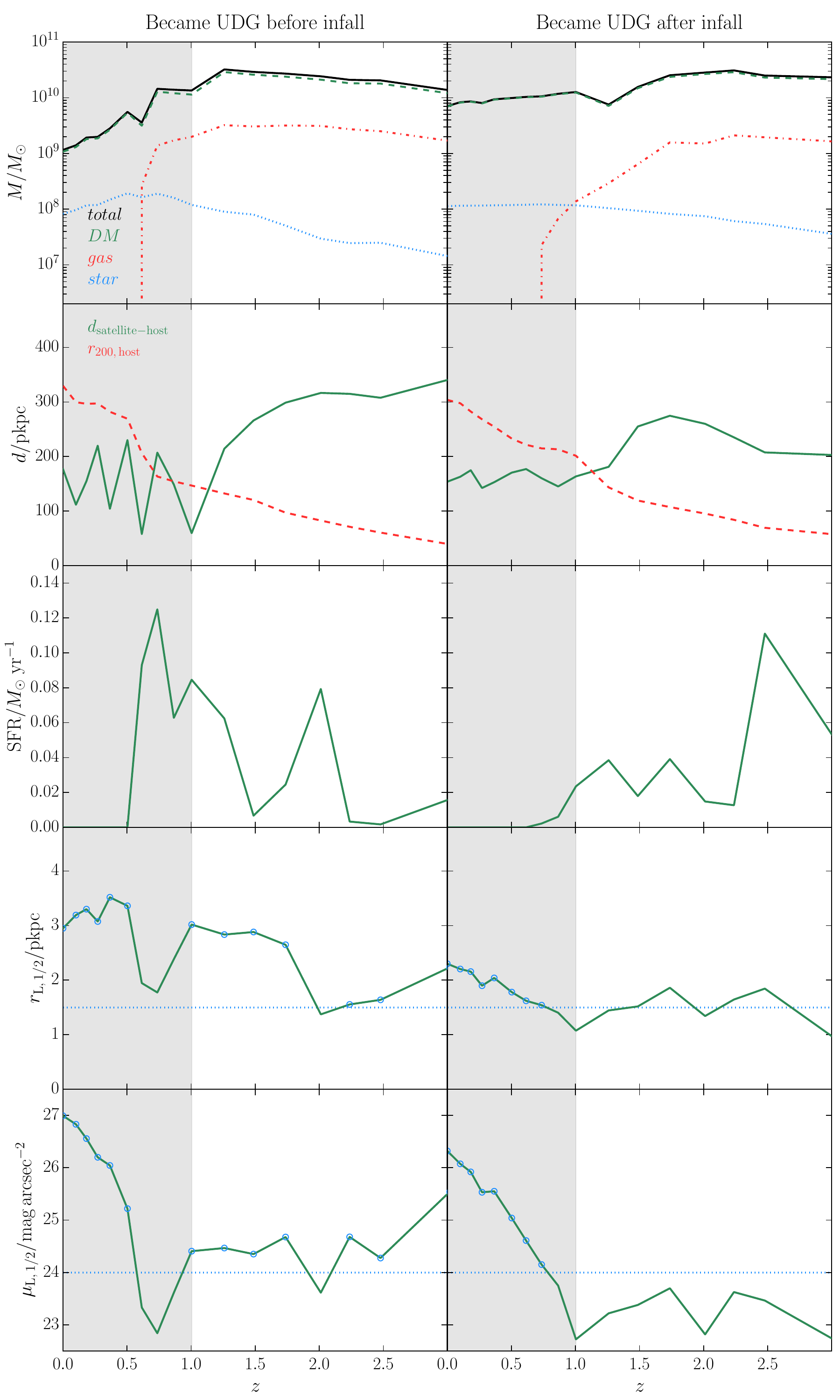}
	\vspace{-0.15cm}
        \caption{Evolution histories of two satellite UDGs representing two different formation mechanisms. The left column shows a satellite that became a UDG before infall, while the right column illustrates one that only became a UDG after infall. The first infall time and the snapshots afterwards (i.e., $z \le 1$) are marked by gray shaded regions. From top to bottom, we present the evolution of the masses of different components, the distance to the host galaxy and the host's virial radius, the star formation rate, the projected half-light radius $r_\mathrm{L,\,1/2}$, and mean surface brightness $\mu_\mathrm{L,\,1/2}$ within this radius. Blue horizontal dotted lines indicate the definition of UDGs (i.e., $r_\mathrm{L,\,1/2} \geq 1.5~\mathrm{pkpc}$, $\mu_\mathrm{L,\,1/2} \geq 24~\mathrm{mag}\,\mathrm{arcsec}^{-2}$), and blue circles mark the moments the satellites are identified as UDGs. Both of these satellites fall into their halo halos at $z=1$, by which time the left case was already a UDG and the right one was not, and after infall, they both went through tidal disruption, causing mass loss and size expansion – interestingly, the left one had a brief star formation boost and temporarily became a normal galaxy, but soon quenched and became dominated by tidal effects. }
	
	\label{fig:fig10}
\end{figure*}

Based on this picture, we speculate that the reason the EAGLE field UDGs exhibit higher spins of stellar and star-forming gas components, while having similar spins of dark matter and total gas, is that they may have a larger fraction of recycled gas. With higher angular momentum in the star-forming gas, they tend to have more gas and more recent star formation activities at their outskirts. 
To test our speculation, we compare the median profile of star formation time, color, and stellar metallicity between UDGs and normal galaxies in Fig. \ref{fig:fig9}. The results show that UDGs have a younger and bluer stellar population at the outskirts, which is consistent with the consequence of the higher spins of recently recycled star-forming gas. Meanwhile, UDGs have a flatter median stellar metallicity profile compared to normal galaxies, aligning with the picture in the literature \citep[e.g.,][]{Angles-Alcazar2014, Ma2017, Grand2019} that galaxies with strong gas outflows tend to redistribute metals from the central regions to the outskirts, resulting in lower central metallicity and a shallower metallicity gradient. 
Note that although this mechanism can also be attributed to feedback-driven gas outflows, it differs from the scenario of dark matter core formation and stellar expansion as tested in Fig. \ref{fig:fig7}.

\subsection{The formation of satellite UDGs}

As noted in previous studies \citep[e.g.,][]{Roman2017b, Chan2018, Liao2019, Jiang2019}, satellite UDGs may have two origins: (i) they were already UDGs before falling into the host halo (i.e., equivalent to the field UDGs); (ii) they become UDGs through tidal effects after infall. In this subsection, we investigate whether this picture applies to EAGLE and if so, we examine whether the fractions of each origin in EAGLE are similar to those in other simulations.

To do this, we trace the evolutionary history (up to $z=3$) of 72 satellite UDGs in EAGLE using their merger trees.\footnote{Among all 73 satellite UDGs, there is one UDG that can only be continuously traced back to $z = 1$, which is insufficient for our study, and therefore is excluded from the following analysis.} Similar to other simulations \citep{Jiang2019,Liao2019}, the formation mechanisms of EAGLE satellite UDGs can also be categorized into the aforementioned origins.

In Fig. \ref{fig:fig10}, we present two case studies corresponding to the two origins, with the left column illustrating the first origin and the right column illustrating the second. Both progenitors of these UDGs fall into their host halos at $z \sim 1$ and soon begin to lose mass. In the figure, we use gray shaded regions to mark the redshift range after the galaxy first entered the virial radius of its host and blue circles to mark snapshots when the galaxy is a UDG. Following \citet{Liao2019}, we define a UDG using the projected half-luminosity radius, $r_\mathrm{L,\,1/2}$ and the mean $g$-band surface brightness inside, $\mu_\mathrm{L,\,1/2}$, instead of the fitted $r_\mathrm{e}$ and $\mu_0$. This choice is made because numerical resolution and tidal effects limit our ability to generate robust fits throughout the satellite history during $z=0\,$--$\,3$.

The case in the left column was already a UDG before entering its host halo's virial radius, with $r_\mathrm{L,\,1/2} = 3.02~\mathrm{pkpc}$ and $\mu_\mathrm{L,\,1/2} = 24.41~\mathrm{mag}\,\mathrm{arcsec^{-2}}$. In contrast, the right case entered as a normal galaxy, with $r_\mathrm{L,\,1/2} = 1.07~\mathrm{pkpc}$ and $\mu_\mathrm{L,\,1/2} = 22.72~\mathrm{mag}$ $\mathrm{arcsec^{-2}}$, and only became a UDG after being puffed up due to tidal effects. Interestingly, the former case, which experienced a temporary boost in star formation rate after falling in, almost doubled its stellar mass and became bright enough in the center to be identified as a normal galaxy. However, it soon suffered from even more aggressive tidal effects, losing $91.44\%$ of total mass compared to the moment of infall and $57.19\%$ of its stars compared to its peak stellar mass. Ultimately, it turned into a UDG again, with $r_\mathrm{L,\,1/2} = 2.95~\mathrm{pkpc}$ and $\mu_\mathrm{L,\,1/2} = 27.00~\mathrm{mag}$ $\mathrm{arcsec^{-2}}$ at $z=0$. This confirms that tidal effects play a crucial role in extending satellite sizes and suggests that multiple factors may impact these sizes.

To quantify the number of satellite UDGs belonging to the two origins, we define the `infall' time as the first snapshot when a galaxy enters its host halo's virial radius, and then categorize a UDG as belonging to the field origin case if it has ever been identified as a UDG in any of the five snapshots before infall. We find that 43 of them (i.e., $\sim 60\%$) were UDGs before infall; this fraction is close to the findings of \citet{Liao2019} and \citet{Jiang2019} ($\sim 55\%$ and $50\%$, respectively). The remaining 29 ($\sim 40 \%$) have a tidal origin.

\section{Summary} \label{sec:conclusion}

In this paper, we utilize the highest-resolution EAGLE run, Recal-L025N0752, to study the properties and formation of UDGs. We identify a total of 181 UDGs from this EAGLE run and compare them with 529 normal dwarf galaxies within the same $g$-band absolute magnitude range $M_{g} \in [-18, -12]$. The EAGLE UDGs exhibit properties that are consistent with observational results, including sizes, central surface brightnesses, stellar masses, S\'{e}rsic indices, and the satellite abundance – host mass relation. Our main findings can be summarized as follows.

(i) The total masses of EAGLE UDGs range from ${\sim}5\times 10^{8}~M_{\odot}$ to ${\sim 2}\times 10^{11}~M_{\odot}$, indicating that they are dwarf galaxies rather than failed $L_\star$ galaxies.

(ii) The comparison of UDG and normal galaxy properties suggests that UDGs are not a distinct population but rather a subset of dwarf galaxies, as their properties generally form a continuous distribution with those of normal dwarf galaxies.

(iii) For field UDGs, we compare them with normal galaxies in terms of the distributions of their halo spin parameters and the evolution of density profiles. Unlike the case in previous simulations, their extended sizes are not due to residing in high-spin halos or stellar expansion caused by supernovae-driven outflows; instead, EAGLE field UDGs exhibit higher spins in star-forming gas and stellar components, and they have formed more new stars at larger radii in recent times. We hypothesize that these can originate from dwarf galaxies containing a higher fraction of star-forming gas that undergoes galactic fountains, acquires angular momentum via mixing with the CGM, and settles in outer radii. This gas eventually forms stars with higher spins, contributing to more extended stellar distributions.

(iv) For satellite UDGs, similar to other simulations, ${\sim} 60 \%$ of them were already UDGs before falling into the host galaxy, while the remaining ${\sim} 40\%$ were normal galaxies prior to infall and only transformed into UDGs due to tidal effects.

Our results indicate another possible formation channel (i.e., galactic fountain) for field galaxies with extended sizes, which can be examined with different simulations in the future, especially in those with more frequent snapshot outputs. 
Meanwhile, we confirm that tidal effects are important for UDG formation in dense environments, with a rate comparable to those reported in the literature. 

\begin{acknowledgments}
We thank the anonymous referee for a very constructive and helpful report which greatly improves our manuscript. 
We thank María Luisa Buzzo, Duncan Forbes, Lucas Valenzuela, and Hang Yang for useful comments, and Kyle Oman, John Helly for sharing \textsc{eagleSqlTools}. We acknowledge the support from the National Key Program for Science and Technology Research and Development of China (2023YFB3002500), the National Natural Science Foundation of China (NSFC) grant (No. 12588202), and the China Manned Space Program with grant No. CMS-CSST-2025-A03 and No. CMS-CSST-2025-A09. HZ acknowledges support from the China Scholarships Council (No. 202104910325). SL acknowledges the support by the NSFC grant (No. 12473015). FJ acknowledges the support by NSFC (No. 12473007) and Beijing Natural Science Foundation (QY23018). This work used the DiRAC@Durham facility managed by the Institute for Computational Cosmology on behalf of the STFC DiRAC HPC Facility (www.dirac.ac.uk). The equipment was funded by BEIS capital funding via STFC capital grants ST/K00042X/1, ST/P002293/1, ST/R002371/1 and ST/S002502/1, Durham University and STFC operations grant ST/R000832/1. DiRAC is part of the UK National e-Infrastructure.

\end{acknowledgments}

{\it Data availability:} The EAGLE simulation data are publicly available at \href{https://icc.dur.ac.uk/Eagle/database.php}{icc.dur.ac.uk/Eagle/database.php} \citep{McAlpine2016, TheEAGLEteam2017}. 

{\it Software:} \textsc{matplotlib} \citep{Hunter2007}, \textsc{numpy} \citep{Harris2020}, \textsc{scipy} \citep{Virtanen2020}.

\bibliography{main}{}
\bibliographystyle{aasjournal}

\end{document}